\documentstyle[prl,aps,epsfig]{revtex}   
\draft

\begin{document}
\title{Destruction of 
Bose-Einstein condensate by strong interactions}

\author{Mariusz Gajda, Magdalena A. Za{\l}uska-Kotur, and Jan Mostowski}
\address{Instytut Fizyki PAN \& College of Science,\\
Aleja Lotnik{\'o}w 32/46, 02-668 Warszawa, Poland}
\maketitle

\begin{abstract}
We study exactly soluble system of trapped bosonic 
particles interacting by a model harmonic
forces. The model allows for detailed examination of the order
parameter (condensate wave function) as well as concept of the off-diagonal
and diagonal order. 
We analyze the effect of interactions on
the condensate and show that sufficiently strong interactions,
attractive or repulsive,  lead to 
destruction of the condensate. In the thermodynamic limit this destruction has
a critical character. It is shown that the existence of the coherent state
of bosons is related to existence 
of two  length scales determined by  
one- and two-particle reduced density matrices. The condensate can exist only
if the two length scales are of the same order. Interactions, both repulsive
and attractive,  change their relative size which may lead to destruction
of coherence in the system and depletion of the condensate. We suggest
that this scenario is model independent. 
\end{abstract}

\pacs{PACS number(s): 03.75.Fi, 05.30.Fk}

\section{Introduction}

Recent  advances in the trapping and cooling techniques which led to the
achievement of the Bose-Einstein condensation of dilute gases have renewed
interest in various aspects of many body theory. A cloud of weakly interacting
trapped atoms is an ideal system for which various aspects of many body theory
can be tested and verified. The ideal bosonic gas undergoes the Bose-Einstein
condensation if the phase-space density exceeds one. This phenomenon manifests
itself by the macroscopic occupation of the single particle ground state. In
the case of an interacting system the condensate wave function can be defined
by spectral decomposition of the one-body reduced density matrix.  This
decomposition is closely related to the off-diagonal long range order
\cite{Onsager,Landau,Yang,Kac} and existence of the order parameter i.e. the
`classical' field with given amplitude and phase commonly used in the theory of
superfluidity \cite{Dalfovo,Griffin}. Realization of decomposition procedure is
practically impossible because it requires a full solution of the many-body
problem.  Instead mean-field approaches are commonly used. The basic idea for a
mean-field description of the dilute, weakly interacting Bose gas below
transition temperature was introduced by Bogoliubov \cite{Bogolubov}. Most of
the results for the interacting Bose-Einstein condensate are obtained within
the Bogoliubov theory which in many cases provides a reliable quantitative
description of the quantum Bose gas. On the contrary, the rigorous treatment is
possible only for  model interactions.

There are only few exactly soluble models of quantum systems where the
interactions between atoms is chosen in the form allowing for the exact
analytic solution. These are: (i) the one-dimensional model of impenetrable
bosons introduced by Girardeau \cite{Girardeau}, (ii) its contact potential
version formulated by Lieb \cite{Lieb}; (iii) the model of particles
interacting by harmonic forces
\cite{Calogero,Bialynicki,Lemmens}. Although in the first two cases
the formal solution is given but in practice the problem is still quite
complicated and quantitative calculations can be done for a very small number
of particles only \cite{Rzazewski}. The latter case seems to be much simpler
because, as it has been shown in \cite{Magda}, it can be reduced to the problem
of noninteracting particles in a harmonic trap. Therefore in the following we
are going to examine, within this exactly soluble model, various concepts 
related to the interacting Bose-Einstein condensate.

This paper is organized as follows. In Sec.\ II we present exact results for
harmonically interacting bosons trapped within harmonic potential \cite{Magda}.
In  Sec.\ III we find the analytic expression for the order
parameter and study the effect of quantum depletion of the condensate as well
as quantum fluctuations at zero temperature. In Sec.\ III we analyse the
off-diagonal and diagonal order and show how off-diagonal long range order
disappears when interactions become very strong.  We finish in Sec.\ IV with
some concluding remarks.

\section{Diagonalization of the Hamiltonian}

In our previous paper \cite{Magda} we have shown the algebraic method of
diagonalization of the Hamiltonian describing a system of many particles
interacting via harmonic forces. The system under consideration 
consists of many particles confined by an external harmonic potential
interacting by harmonic forces, i.e. two body interaction potential has the
form:
\begin{equation}
V({\bf x}_i - {\bf x}_j) = \frac{\sigma}{2}\Omega^2 ({\bf x}_i - {\bf
x}_j)^2, 
\end{equation}
where $\Omega$ defines the interaction strength and $\sigma = +1$ signifies the
attractive interaction of particles placed at positions ${\bf x}_i$ and ${\bf
x}_j$ whereas $\sigma =-1$ -- corresponds to repulsive interactions. 
The total Hamiltonian of the $N$-particle system has therefore the following
form: 
\begin{equation} 
H = \sum_{i=1}^N \frac{1}{2}({\bf p}_i^2 + {\bf x}_i^2)
+ \sum_{i<j} V({\bf x}_i - {\bf x}_j).  
\label{a0}
\end{equation}

Let us first recall some exact results of \cite{Magda}. For the sake of
simplicity we denote the set of all particle positions vectors 
by ${\bf X}_N = ({\bf x}_1, \ldots, {\bf x}_N)$.  The Hamiltonian which is a
quadratic form of positions and momenta of particles can be easily
diagonalized if one introduces collective variables:
\begin{equation}
{\bf X}^c_N = {\cal Q_N}\, {\bf X}_N,
\end{equation}
where ${\bf X}^c_N = ({\bf x}^c_1, \ldots, {\bf x}^c_N)$ 
and the matrix ${\cal Q}_N = \{ q^N_{ij} \}$ is orthogonal. 
One of these collective variables namely the center of mass of $N$-particle
system plays a particularly important role:
\begin{equation}
{\bf x}^c_N = \frac{1}{\sqrt{N}} \sum_{i=1}^N {\bf x}_i.
\end{equation}  
The choice of $N-1$ remaining variables ${\bf X}^{c}_{N-1} =
({\bf x}^c_1, \ldots, {\bf x}^c_{N-1})$ is not unique
but this does not lead to any physical implications. In particular:
\begin{equation}
({\bf X}^c_{N-1})^2=\sum_{i=1}^{N-1} ({\bf x}^c_i)^2 = 
\sum_{i=1}^{N} ({\bf x}_i)^2 - ({\bf x}^c_N)^2.
\end{equation}
In the following we are going to use a similar notation for description of a
subsystem of $s$-particles, $s=1, \ldots,N$.

The transformation defined above brings the Hamiltonian to the diagonal form
and its eigenenergies can be easily found. However, while determining a
spectrum one must also take into account the proper symmetry of a total 
wave function. In the case of bosonic particles ($N > 2$) the allowed energies
are: 
\begin{equation}  
\label{spec}
E=\left( \frac{3}{2} +m \right) + \left(\frac{3}{2} (N-1)+n \right)\omega,
\end{equation}
where $m=0,1,2\ldots$, $n=0,2,3\ldots$ and $\omega = \sqrt{1+ \sigma
N\Omega^2}$. 
The first term describes excitations of the center of mass -- d-dimensional
harmonic oscillator of frequency equal to one. The second term in the 
Eq.(\ref{spec}) corresponds to excitations of $N-1$ relative degrees of
freedom. The frequency $\omega$ characterizes the effective potential
experienced by $N-1$ collective modes. 

Let us observe that $\omega = 1$ 
corresponds to the noninteracting case, the repulsive interactions give $0<
\omega < 1$ while  attractive forces lead to $\omega > 1$. Moreover, very small
values of $\omega \approx 0$ signify very strong repulsion which almost
destabilizes the whole system. It is very convenient to parameterize $\omega$
by an exponent $\kappa$ defined in the following way:
\begin{equation}
\label{kapa}
\omega = N^{\kappa}.
\end{equation}
This exponent can be related to the actual strength of the interaction. In
fact, for  weakly interacting gas ($\omega \approx 1$) we  obtain very small
values of this parameter: $\kappa \approx 0$, while for strong interactions
($\omega \approx 0$ -- repulsion,  $\omega \gg 1$ -- attraction) we have
$|\kappa| \gg 1$.  Moreover, $\kappa$ is positive in the case of attraction
while it is negative for repulsion. Let us add at this point that in 
realistic situations of  short-range interparticle interactions, large
Bose-Einstein condensates can exist only for  repulsive forces. In the
case of attraction the size of the trapped condensate is limited to about
1500 atoms \cite{Dalfovo}. In our oscillatory model the forces between
particles are negligible at small distances, therefore the model leads to the
condensation (in the thermodynamic limit) in both attractive and repulsive
case.  

The ground state of the system is the following:
\begin{equation}
\label{wfun}
\Psi({\bf X}_N)=\Phi_0(\sqrt{\omega} {\bf X}^c_{N-1}) \Phi_0({\bf x}^c_N),
\end{equation}
where $({\bf X}^c_{N-1},{\bf x}^c_N) = {\cal Q}_N {\bf X}_N$ and 
the function $\Phi_0(\sqrt{\omega} {\bf X}^c_{N-1})$ corresponds to the ground 
state of a system of $N-1$ independent quasi-particles (in $d$ spatial
dimensions) 
interacting with an external potential of the harmonic oscillator of frequency
$\omega$: 
\begin{equation}
\label{phirel}
\Phi_0(\sqrt{\omega} {\bf X}^c_{N-1}) = 
\left( \frac{\omega}{\pi} \right)^{d(N-1)/4} 
{\rm exp}\left[-\omega ({\bf X}^c_{N-1})^2/2\right],
\end{equation}
and $\Phi_0({\bf x}^c_N)$ is the ground state of the single particle (center of
mass) trapped into harmonic potential:
\begin{equation}
\label{phicm}
\Phi_0({\bf x}^c_N) = 
\left( \frac{1}{\pi} \right)^{d/4} 
{\rm exp}\left[-({\bf x}^c_N)^2/2\right].
\end{equation}
Construction of excited eigenstates is difficult because it is not easy to
impose the desired symmetry on the wave function. This procedure was describe in
details in \cite{Magda}.

\section{Order parameter and quantum depletion}
If the energy of the system (or equivalently the temperature) is sufficiently
small we expect that the system forms a Bose-Einstein condensate.  The BEC of
the ideal gas manifests itself by a macroscopic occupation of the single
particle ground state. In the case of interacting system 
it is not obvious what is this particular state which is
`macroscopically occupied'. The identification of the macroscopically occupied
quantum state is equivalent to the definition of the order parameter -- the
single particle wave function which is inherently related to the Bose
condensation. The condensate subsystem can be then quite accurately described
by the $N_0$-fold product of the order parameter, where $N_0 \simeq {\cal
O}(N)$ is the occupation 
of the condensate. 

In the following we use our model to demonstrate how to define the order
parameter, occupation of the condensate, and its fluctuations. 
At zero temperature the system
is in the ground state and one might naively expect that it is totally Bose
condensated. However, the ground state of the N-particle bosonic system is not
equivalent to the Bose-Einstein condensate. Interactions can significantly
deplete the condensate. We are going to show this effect in the most
spectacular but also in relatively simple case of the zero temperature. 

Let us now define the hierarchy of the reduced $s$-particle density matrices
which can be conventionally obtained by averaging the density matrix of the
total system of $N$ particles over the degrees of freedom of $N-s$ remaining
particles. For a given $N$-particle quantum state $\Psi({\bf X}_N)$ the
corresponding $s$-particle reduced density matrix $\rho_s({\bf X}_s; {\bf
Y}_s)$ is defined by: 
\begin{equation}
\label{rho1}
\rho_s({\bf X}_s; {\bf Y}_s)=
\int {\rm d}{\bf R}_{N-s} \Psi^*({\bf X}_s,{\bf R}_{N-s})
\Psi({\bf Y}_s,{\bf R}_{N-s}).
\end{equation}
We use previously defined shorthand notation for vectors in the configuration 
space of $s$-particles, for example ${\bf X}_N = ({\bf X}_s, {\bf R}_{N-s})$
and ${\bf X}_s = ({\bf x}_1, \ldots, {\bf x}_s)$.
The reduced density matrix describes 
the subsystem of $s$-particles and can be directly related to different
measurement processes. For the statistical description of the system
one should first of all define the statistical density matrix by averaging
all $N$-particle density matrices with the appropriate statistical weights
depending on the ensemble. In general it is quite a complicated task
but at zero temperature there is only one quantum state of the system and
no statistical averaging is necessary.

The total wave function (or density matrix) carries all the
information about the system. In real experiments however one does not probe
simultaneously all the particles. Typical detection scheme consists on the
measurement of one or at most few particles at a given time. In other words a
single measurement process is reduced to 
a subsystem of small number of particles. Such subsystems are
described by reduced density matrices. In the considered here case of
zero temperature the $s$-particle
density matrix can be brought to the following form:
\begin{equation}
\label{s-part}
\rho_s({\bf X}_s; {\bf Y}_s) = \rho^{CM}({\bf x}^c_s, {\bf y}^c_s)
\Phi_0(\sqrt{\omega} {\bf X}^c_{s-1}) \Phi_0(\sqrt{\omega} {\bf Y}^c_{s-1}).
\end{equation}
The functions $\Phi_0(\sqrt{\omega}{\bf X}^c_{s-1})$ describes the ground state
of $s-1$ quasi-particles:
\begin{equation}
\Phi_0(\sqrt{\omega} {\bf X}^c_{s-1}) = 
\left( \frac{\omega}{\pi} \right)^{d(s-1)/4} 
{\rm exp}\left[-\omega ({\bf X}^c_{s-1})^2/2\right],
\end{equation}
while $\rho^{CM}({\bf x}^c_s, {\bf y}^c_s)$ corresponds
to the density matrix of center of mass of the subsystem:
\begin{equation}
\label{reduced}
\rho^{CM}({\bf x}^c_s, {\bf y}^c_s) = \left( \frac{{\omega}_s}{\pi}
\right)^{d/2} 
{\rm exp}\left\{ 
      -\frac{1}{2}
      \left(\omega_s + \frac{\delta_s}{2}\right) 
      \left[({\bf x}^c_s)^2 + ({\bf y}^c_s)^2 \right]+
      \frac{\delta_s}{2} \, {\bf x}^c_s \,{\bf y}^c_s
	\right\}.
\end{equation} 
The $s$-particles collective coordinates are defined in the familiar way:
$({\bf X}^c_{s-1}, {\bf x}^c_s) = {\cal Q}_s \,{\bf X}_s$, 
$({\bf Y}^c_{s-1},{\bf y}^c_s)= {\cal Q}_s \, {\bf Y}_s$ and frequencies
$\omega_s$, $\delta_s$ as well as auxiliary parameter $\gamma_s$ are:
\begin{eqnarray}
\gamma_s &=& 1 -\frac{s(1-\omega)}{N},\\
\omega_s &=& \frac{\omega}{\gamma_s}, \\
\delta_s &=& \left( \frac{1-\omega}{N} \right)^2 \frac{s(N-s)}{\gamma_s}.
\end{eqnarray} 

Having defined the $s$-particle matrices we are ready now to analyze the nature
of the Bose-Einstein condensation of the interacting system and to discuss the
meaning of the order parameter. To this end we write the density matrix
Eq.(\ref{s-part}) in the diagonal form:  
\begin{equation}
\label{spectral}
\rho_s({\bf X}_s; {\bf Y}_s) = 
\sum_{\bf n} \lambda^{(s)}_{\bf n} \phi^{(s)}_{\bf n}({\bf X}_s)\, 
\phi^{(s)}_{\bf n}({\bf Y}_s). 
\end{equation}
The function $\phi^{(s)}_{\bf n}({\bf X}_s)$ can be treated as the wave
function of the $s$-particle subsystem:
\begin{equation}
\label{cm}
\phi^{(s)}_{\bf n}({\bf X}_s)=\Phi_0(\sqrt{\omega}{\bf X}^c_{s-1})
\Phi_{\bf n}(\sqrt{\alpha_s} {\bf x}^c_s),
\end{equation}
where $\Phi_0(\sqrt{\omega}{\bf X}^c_{s-1})$ is the ground state wave function
of the relative degrees of freedom. This function corresponds to the ground
state of $s-1$ noninteracting quasi-particles (in d-spatial dimensions) subject
to the external harmonic potential of frequency $\omega$. The second part of
the Eq.(\ref{cm}) describes states of the center of mass of $s$-particles;
$\Phi_{\bf n}$ is simply the d-dimensional harmonic oscillator wave function
corresponding to the effective center of mass frequency $\alpha_s$. Quantum
numbers ${\bf n}=(n_1, \ldots, n_d)$ label different states of the center of
mass while $n=n_1+ \ldots +n_d$ corresponds to the energy of the given state.
The effective center of mass frequency $\alpha_s$ is:
\begin{equation}
\alpha_s = \left[ \omega_s (\omega_s + \delta_s) \right]^{1/2}.
\end{equation}
It is interesting to observe that all the frequencies of the relative motion
of the $s$-particles subsystem are exactly the same as the frequencies of the
relative motion of the whole system, i.e. equal to $\omega$. On the other hand
the center of mass oscillation frequency of the 
subsystem is neither equal to $\omega$ nor to 1 (trap frequency). This
collective degree of freedom couples to the center of mass of $N-s$ remaining
particles what leads to some energy shift.
Finally, the eigenvalues $\lambda^{(s)}_n$ of $\rho_s$ are equal to
the occupation probabilities of a given $s$-particle state: 
\begin{equation}
\label{lambda}
\lambda^{(s)}_n=\left(\frac{\omega_s}{\alpha_s}\right)^{d/2}
\left(\frac{2 \sqrt{\omega_s \alpha_s}}{\omega_s+\alpha_s}\right)^d
\left(\frac{\alpha_s-\omega_s}{\alpha_s+\omega_s}\right)^n.
\end{equation}
It follows from the normalization condition for the density matrix that:
$\sum_{\bf n} \lambda^{(s)}_n =1$.

The spectral decomposition of the reduced one-particle density matrix gives
natural single-particle states $\phi^{(1)}_{\bf n}({\bf x})$. These
states are crucial for the definition of the condensate wave function (order
parameter). It can be seen from Eq.(\ref{lambda}) that if $N$ goes to
infinity (thermodynamic limit) with fixed value of the interaction strength
$\kappa$ the lowest eigenvalue $\lambda^{(1)}_0$ dominates the others 
and in the limit of weak interactions we get: 
\begin{eqnarray}
\lambda^{(1)}_0       &\simeq& 1,\\
\lambda^{(1)}_{\bf n} &\simeq& \left(\frac{(1-\omega)^2}{4N}\right)^n,
\phantom{123}{\rm if}\phantom{123} n \neq 0.
\end{eqnarray}
This behavior signifies nothing else but the onset of the Bose-Einstein
condensation. The single particle density matrix becomes very close to the pure
state because with quite good accuracy it can be approximated by $\rho_1( {\bf
x}, {\bf y}) \approx \phi^{(1)}_0({\bf x})\, \phi^{(1)}_0({\bf y})$. This
particular single-particle ground state $\phi^{(1)}_0({\bf x})$ is usually
called the order parameter. The $N$-particle wave function can be quite
accurately approximated by the $N$-fold product of the order parameter. 

Our analytic formula allows to study quantitatively the role of interactions on
the Bose-Einstein condensate.  On the basis of the discussion it is obvious
that the average occupation of the condensate becomes:
\begin{equation}
\label{n0}
\langle N_0 \rangle = N \int {\rm d}{\bf x} {\rm d}{\bf y} \
\phi^{(1)}_0 ({\bf x}) \ \rho_1({\bf x}, {\bf y}) \
{\phi^{(1)}_0}({\bf y}) = N \lambda^{(1)}_{0}.
\end{equation}
In the case of the ideal gas at zero temperature the above equation gives, of
course, $\langle N_0 \rangle = N$; all particles occupy the single particle
ground state. For studied here model of the interacting system it is
more convenient to use as independent variables the pair of $\kappa$ and $N$
rather than $\omega$ and $N$. The exponent $\kappa$ occurs to be inherently
related to the oscillatory interactions because when using this parameter we
discover some universal features. For a fixed number of particles, if the
interaction strength 
$|\kappa| \sim \log\omega$ grows, the occupation of the condensate decreases. This
behavior is presented in Fig. 1 where we show the mean occupation of the
condensate versus the exponent $\kappa= \log\omega/ \log N$ for different
values of particle number $N$ in three spatial dimensions ($d=3$). The values
of $\kappa$ less than zero signify repulsive interactions while $\kappa > 0$
corresponds to attraction. One can easily see that if the interaction becomes
strong ($|\kappa| \simeq 1$) the condensate is almost totally depleted. All
curves presented in the figure tend to an universal curve if the number
of particles increases.  When $N$ increases to infinity with $\kappa$ being
constant then our expression for the occupation of the condensate has the form:
\begin{equation}
\label{ocup}
\frac{\langle N_0 \rangle}{N}=
\left(\frac{2}{1+\sqrt{N^{\kappa-1}+N^{-(\kappa+1)}+1}}\right)^d.
\end{equation}
The above formula, valid in the thermodynamic limit,  gives an universal
critical behavior. It exhibits no depletion ($N_0 = N$) for
$|\kappa| <1$ followed  by an abrupt jump and total destruction of the
condensate  ($N_0=0$)  for $|\kappa|>1$.
\begin{figure}
   \begin{center}
   \epsfxsize 7.5cm
   \epsffile{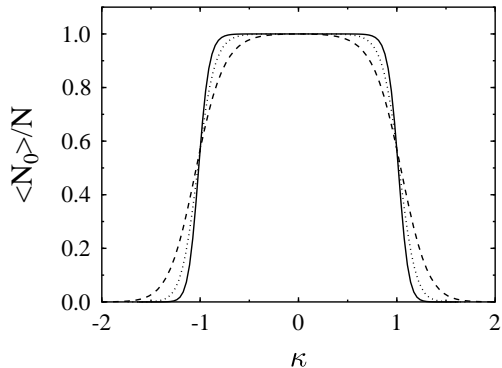}
   \end{center}
   \caption{
            Mean occupation of the condensate plotted as a function
            of the parameter $\kappa = \log\omega/\log N$ for different 
            number of
            particles; $N=10^3$ -- dashed line; $N=10^5$ -- dotted line,
            and $N=10^8$ -- full line.
            }
   \label{fig1}
\end{figure}     

The effect of quantum depletion of the trapped atomic condensate with a short
range interactions, for the realistic experimental parameters, has been
estimated to be of the order of $1\%$ \cite{Dalfovo}. This is opposite to the
case of superfluid helium where this effect accounts for depletion as big as
more  than $90\%$ \cite{Huang}. Our model exhibits very interesting feature. It
shows that in large $N$ limit the quantum effects are almost negligible or
totally destroy the condensate depending on the value of the interaction
strength. At this point it is not clear if this is an unique feature of
our model or if it is more general result.

The 2-particle reduced density matrix allows to find a joint probability of
finding a particle in a given single particle state provided that
another particle is also in some given state. In particular we have: 
\begin{equation}
\langle N_0 (N_0 -1) \rangle = N(N-1)
\int {\rm d}{\bf X}_2 {\rm d}{\bf Y}_2 \ \left[\phi^{(1)}_{0}({\bf x}_1)
\phi^{(1)}_{0}({\bf x}_2)\right] \ \rho_2({\bf X}_2, {\bf Y}_2) \
\left[{\phi^{(1)}_{0}}({\bf y}_2) {\phi^{(1)}_{0}}({\bf y}_1)\right].
\end{equation}
Simple integration gives:
\begin{equation}
\langle N_0 (N_0 -1) \rangle = N(N-1)
\left( \frac{2 \sqrt{\omega \alpha_1}}{\omega + \alpha_1}\right)^d
\left( \frac{2 \sqrt{\omega_2 \alpha_1}}{\omega_2 + \alpha_1} \right)^d 
\left(\frac{\omega_2 +\alpha_1}{\omega_2+\alpha_1+\delta_2}\right)^{d/2}.
\end{equation}
Now we are ready to analyze the fluctuations of the condensate defined as:
\begin{equation}
\label{fluk}
\langle \delta^2 N_0 \rangle = \langle (N_0)^2 \rangle - \langle N_0 \rangle^2.
\end{equation}
These fluctuations are shown in Fig.2. We see that as the interaction
strength grows (at fixed number of particles) the fluctuations start to grow
from zero value for the ideal gas. However, again when the interactions
become so strong that condensate practically disappears ($|\kappa| \simeq 1$)
fluctuations also decrease - as there is no condensate the fluctuations also
die out. The fluctuations are maximal in a region of the critical destruction
of the condensate by quantum effects.
\begin{figure}
   \begin{center}
   \epsfxsize 7.5cm
   \epsffile{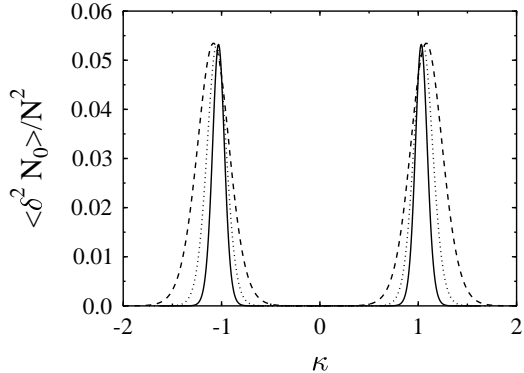}
   \end{center}
   \caption{
            Fluctuations of the condensate plotted as a function
            of the parameter $\kappa = \log\omega/\log N$ for different 
            number of
            particles; $N=10^3$ -- dashed line; $N=10^5$ -- dotted line,
            and $N=10^8$ -- full line.
            }
   \label{fig2}
\end{figure}     

\section{Off-diagonal order and correlations}
In this subsection we are going to study (within our model) the phase and
density correlations for  
the ground state of the interacting system of bosonic particles. In the early
works of Penrose and Onsager \cite{Onsager} it has been shown that phenomenon
of Bose condensation is evidenced by the presence of the off-diagonal long
range order in the one-particle density matrix $\rho_1({\bf x}, {\bf y})$:
\begin{equation}
\label{odlro}
\lim_{|{\bf x}-{\bf y}| \rightarrow \infty} \rho_1({\bf x},{\bf y}) \neq 0.
\end{equation}
The above condition signifies a large scale correlations of the off-diagonal 
elements of the
single-particle matrix.
The limiting value of correlations is called the off-diagonal long range order
parameter. In a finite system this concept cannot be strictly applied as the
reduced density matrix vanishes at large distances. However, for a practical
purposes the off-diagonal long range order can be related to the behavior of
the off-diagonal elements of the density matrix at distances equal to the size
of the system. 

In order to identify long range correlations we should first
analyze characteristic length scales. It is natural to relate the size
of the system to the spatial extension of the single particle
density, i.e. the diagonal elements of the one-particle matrix: 
\begin{equation}
\label{den}
n_1({\bf x}) = N \rho_1({\bf x}, {\bf x})=
N \left(\frac{\omega_1}{\pi}\right)^{d/2} \exp(-\omega_1 {\bf x}^2).
\end{equation}
Therefore, the size of our system is:
\begin{equation}
\ell_1 = \frac{1}{\sqrt{\omega_1}}.
\end{equation}
For large values of $N$ an extension of the system for attractive
interactions approaches the value of one, $\ell_1 \approx 1$, i.e. is governed
by the characteristic length scale of the external potential felt by the center
of mass. Relative degrees of freedom are localized within much smaller
distances $1/\sqrt{\omega} \ll \ell_1$. For repulsive interactions a
spatial extension of the system is related to the length scale of a mean
field potential experienced by relative degrees of freedom, $\ell_1 \approx
1/\sqrt{\omega}$. In this case the magnitude of the center of mass spreading 
is relatively small $1 \ll \ell_1$. These features have serious
implications on behavior of the order parameter and off-diagonal order. Let us
stress that existence of different length scales -- the one related to the
external potential, and another related to the interparticle interactions
is the general
feature of all two-body central forces.
This is not an unique behavior of considered here long range oscillatory model.

The presence of the off-diagonal long range order in the case of inhomogeneous
system of finite spatial extension manifests itself by a large distance
behavior of the following correlation function:
\begin{equation}
\label{g1}
g_1({\bf x}, {\bf y})= \frac{N \rho_1({\bf x},{\bf y})} 
{\sqrt{ n_1({\bf x}) n_1({\bf y})} }.
\end{equation}
Accordingly, we shall refer to notion of the off-diagonal long range order if:
\begin{equation}
\label{odlro2}
\lim_{|{\bf x}-{\bf y}| \rightarrow \ell_1} 
g_1({\bf x},{\bf y}) \neq 0.
\end{equation}
By dividing the correlation function by the square root of single particle
densities we ensure that correlations are not sensitive to the local density of particles. 
Let us observe that off-diagonal correlation function describes phase correlations
in our system.  In the case when the one-particle density 
matrix corresponds to a pure
state $\Psi({\bf x}) = R({\bf x})\exp[{\rm i} S({\bf x})]$ (where $R({\bf x})$
and $S({\bf x})$ are the modulus and phase of the wave function),
then $g_1({\bf x}, {\bf y}) = \exp\{-{\rm i}[S({\bf x}) - S({\bf y})]\}$.
In the context of the Bose-Einstein condensation of trapped
alkali gases such phase correlations in a trapped 
2-dimensional interacting system has been studied recently in Ref.\cite{Gora}.

\begin{figure}
   \begin{center}
   \epsfxsize 7.5cm
   \epsffile{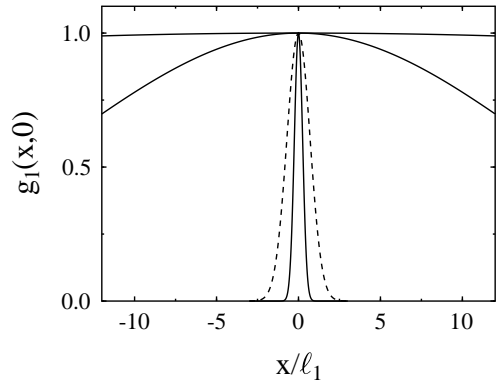}
   \end{center}
   \caption{
            Off-diagonal correlations $g_1(x,0))$ for  a system of $N=10^5$
            particles for different interaction strength
            $\kappa = \log\omega/\log N$. Note that $g_1(x,0)$ does not 
            depend on the
            sign of interaction. The full lines correspond to the following
            values of $|\kappa|$ (from the top to the bottom);  $|\kappa| =
            0.3, \ 0.6, \ 1.3$. The dashed line corresponds to the single
            particle density function $n_1(x)/N$. The distance has been 
            scaled in such a way that single particle density does
            not depend on the strength of interaction.
            }
   \label{fig3}
\end{figure}     

Now we are going to apply the above definition to study the coherence of the ground
state of bosonic particles interacting by harmonic forces. Similarly as 
in the previous section we will investigate the large $N$ limit 
varying the interaction strength $\kappa$. Because the system is inhomogeneous
the correlation function depends on the six variables (in three dimensional
space). To simplify our analysis we fix the position ${\bf y}$ at the trap
center, ${\bf y}=0$.  With this restriction the correlation function depends
only on the distance $x=|{\bf x}|$:  
\begin{equation}
\label{off1}
g_1(x,0) = \exp\left(-\frac{\delta_1}{4} x^2\right),
\end{equation}
what gives for the limit of large distances, $x \sim \ell_1$, the expression for
the off-diagonal order: 
\begin{equation}
\label{off2}
g_1(\ell_1,0) = \exp\left(-\frac{\delta_1}{4 \omega_1} \right).
\end{equation}
In the Fig.3 we show the behavior of the off-diagonal order 
Eq.(\ref{off1}) as a function of distance to the trap center for $N=10^5$
particles and different interaction strength $\kappa$.  We see that for a weak
interaction the system exhibits the off-diagonal long range order, while if
interactions are strong the phase correlations rapidly vanish at distances
smaller than the natural size of the system. For comparison we show
in the Fig.3 the single particle density
$n(x)/N$ (dashed line). This kind of dependence of the off-diagonal
order is consistent with the
behavior of the order parameter (the condensate wave function) discussed
previously. Indeed, the coherence length $L^{(1)}_{off}$ which appears in the
equation Eq.(\ref{off2}) in the large $N$ limit has a form:
\begin{equation}
L^{(1)}_{off} = \left(\frac{\omega_1}{\delta_1}\right)^{1/2} =  \frac{1}{\sqrt{ N^{-\kappa -1} +
N^{\kappa-1}}}.
\end{equation}

Let us notice that the coherence length is governed by the same type 
of $\kappa$-dependence
as the mean occupation of the condensate Eq.(\ref{ocup}). Therefore it is not
surprising that the order parameter and off-diagonal order behaves similarly.
Disappearing of the off-diagonal long range order has exactly the same critical
character as destruction off the order parameter when the interaction strength
exceeds value one, $|\kappa| >1$.  The coherence length in large $N$ limit,
$N \rightarrow \infty$ is:
\begin{eqnarray}
L^{(1)}_{off} &=& \infty, \phantom{12345}{\rm if}\phantom{12}|\kappa| < 1, \\
L^{(1)}_{off} &=& 0, \phantom{123456}{\rm if}\phantom{12}|\kappa| >1.
\end{eqnarray}
So far we have studied the off-diagonal correlations for the one-particle
reduced matrix.  However, a similar analysis can be done for any $s$-particle
density matrix. Without going into unnecessary algebra one can easily estimate
the coherence length for a given $s$-particle subsystem. The form of the
corresponding reduced density matrix (see Eq.(\ref{reduced})) explicitly
suggests that the order of magnitude of the off-diagonal elements at distances
comparable to the size of the system, $x = \ell_1$ can be estimated by:
$\exp[-\delta_s/(4\omega_s)(x/\ell_1)^2]$. 
Therefore, the $s$-order coherence length is:
\begin{equation}
\label{Ls}
L^{(s)}_{off} = \left(\frac{\omega_s}{\delta_s}\right)^{1/2} =  \frac{1}{\sqrt{s( N^{-\kappa -1} +
N^{\kappa-1})}}.
\end{equation}
This expression shows that there is an off-diagonal 
long range coherence on each level of hierarchy of the reduced density matrices
provided that $|\kappa|<1$.

This statement, however, requires some comment. In fact the equation
Eq.(\ref{Ls}) is valid only if  $\lim_{N \rightarrow \infty} (s/N) =0$, i.e. if
we consider the subsystem of finite number of particles (while the total number
of particles is infinite). On the contrary, in the limit $\lim_{N \rightarrow
\infty} (s/N) = \beta >0$ we have:
\begin{equation}
\label{Ls2}
L^{(\beta)}_{off} = \left(\frac{\omega_s}{\delta_s}\right)^{1/2} =  
\frac{1}{\sqrt{\beta(1-\beta)(N^{-\kappa} +N^{\kappa})}} \longrightarrow 0.
\end{equation}
There is no coherence in large subsystems when the number of particles in the
subsystem is of the order of the total number of particles, $s \sim {\cal
O}(N)$. However, as long as the detection scheme is reduced to an observation
of at most few particles simultaneously (what is true for all experiments) the
Bose system at zero temperature can be viewed as described by the coherent
state of many particles. This point of view is justified on every level of the
reduced $s$-particle density matrix provided that $s/N \ll 1$. This kind of
behavior is in a spirit of Yang (\cite{Yang}) conjecture, who suggested that
Bose-Einstein 
condensation is evidenced by the fact that every $s$-particle matrix can be
related to a simple $s$-fold product of the order parameter. In our model of
the interacting condensate the above statement is true for a whole hierarchy
of $s$-particle matrices when $s/N \rightarrow 0$. 
If $s/N \ll 1$ and $|\kappa| < 1$ simple integration gives:
\begin{equation}
\label{prod}
\rho_s({\bf X}_s,{\bf Y}_s) \simeq  \beta^{(s)} 
[\Phi^{(1)}_0({\bf x}_1)\ldots\Phi^{(1)}_0({\bf x}_s)]\
[\Phi^{(1)}_0({\bf y}_1)\ldots\Phi^{(1)}_0({\bf y}_s)],
\end{equation}
where $\beta^{(s)}$ is:
\begin{equation}
\beta^{(s)} =
\left( \frac{2 \sqrt{\omega \alpha_1}}{\omega + \alpha_1}\right)^d
\left( \frac{2 \sqrt{\omega_s \alpha_1}}{\omega_s + \alpha_1} \right)^d 
\left( \frac{\omega_s +\alpha_1}{\omega_s+\alpha_1+\delta_s}\right)^{d/2}
\longrightarrow 1.
\end{equation}
The equation Eq.(\ref{prod}) is not valid if one considers $s$-particle
matrices for $s$ being of the order of the total number of particles, $s \sim
{\cal O}(N)$.
At this point we want to make a comment about the notion of the coherence of
the Bose-Einstein condensate.  Approximate methods assume explicitly that a
mean value of the boson field operator is different than zero in the case of
the Bose-Einstein condensate.  Therefore, the folk wisdom associates the
condensate with the coherent state -- the analogue of the coherent state of the
electromagnetic field. This analogy is of limited value and in fact may be
misleading because the condensate must be in a Fock state in which a mean value
the field operator vanishes. However, there is coherence in the condensate in
the sense that majority of particles is described by the same wave function
with the same phase. The coherence of the condensate manifests itself by the
off-diagonal long range order.  The coherence length of a phase of the
condensate is at least of the same magnitude as the spatial extension of
the system. Moreover this coherence is present also when one considers the
subsystems of any number of particles significantly smaller than a total
particle number.  This facts allows for
substantial simplification of description of many particle system.  When the
system is coherent, i.e. when one can assign exactly the same wave function to
every particle then a behavior of the condensate can be correctly described by
this single-particle function. The above does not apply to measurements based
on simultaneous detection of almost all particles. In such a case the $N$-body
wave function has to be used instead.

The presented analysis of the order parameter and off-diagonal correlations 
exposed a kind of symmetry with respect to change of the interaction sign.
The differences can be seen while studying the diagonal correlations
of the reduced density matrices.  
The diagonal elements of one-particle
matrix are related to single particle density, $n_1({\bf x})=N \rho_1({\bf x},
{\bf x})$ while $\rho_2({\bf x}, {\bf y}; {\bf x}, {\bf y})$ describes a
joint probability of finding a particle at position ${\bf x}$ provided than
another particle is at position ${\bf y}$. The corresponding two-particle
density (if ${\bf x} \neq {\bf y}$) is:
\begin{equation}
n_2({\bf x}, {\bf y}) = N(N-1) \rho_2({\bf x}, {\bf y}; {\bf x}, {\bf y}).
\end{equation}
The above function can be related to two-particle correlation function:
\begin{equation}
g_2({\bf x}, {\bf y}) = \frac{n_2({\bf x},{\bf y}) - 
n_1({\bf x}) n_1({\bf y})}{n_1({\bf x}) n_1({\bf y})}.
\end{equation}
If two-particle density can be written as a product of corresponding
one-particle ones then there are no diagonal (density) correlations, i.e.
$g_2({\bf x},{\bf y}) = 0$. Positive values of $g_2$ signify some clustering
effect while negative values correspond to some effective repulsion. Note that
$g_2$ can be in principle infinitely large (extremely strong attraction) but its
minimal value is limited by $-1$. This value corresponds to extremely strong
repulsion when particles tend to avoid each other. 

In the case of studied here model, all density functions are
inhomogeneous. Therefore, for the sake of simplicity we again fix one of the
position vectors at the trap center, i.e. we set ${\bf y} =0$. Then the
correlation function $g_2$ depends only on the distance $x=|{\bf x}|$:
\begin{equation}
g_2(x,0) = \frac{N-1}{N} 
\left( \frac{ \omega \omega_2}{\omega_1^2} \right)^{d/2} 
\exp\left[ -\frac{1}{2} \left(\omega+\omega_2-2 \omega_1\right) x^2\right]-1.
\end{equation}
The above equation gives the characteristic length scales of two-particle 
diagonal correlation: 
\begin{equation}
L^{(2)}_{diag} = \left(\frac{\omega_1}{\omega+\omega_2-2\omega_1}\right)^{1/2}.
\end{equation}
\begin{figure}
   \begin{center}
   \epsfxsize 7.5cm
   \epsffile{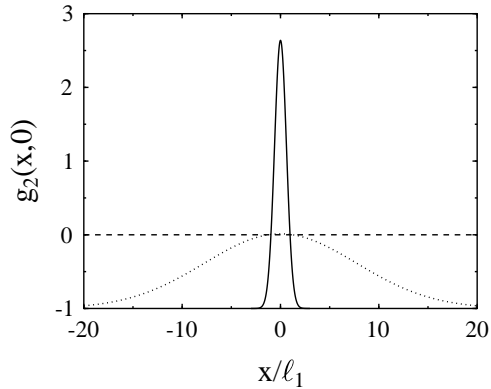}
   \end{center}
   \caption{
            Diagonal correlation $g_2(x,0))$ for  a system of $N=10^5$
            particles for different interaction strength
            $\kappa = \log\omega/\log N$; $\kappa =  0.4$ -- dashed line;
            $\kappa = 0.8$ -- dotted line; $\kappa = 1.1$ -- full line. 
            }
   \label{fig4}
\end{figure}     
Let us first consider the attractive forces.
In this case the diagonal correlation function strongly 
depends on the magnitude of the interaction strength $|\kappa|$.
For weak interactions, $0< \kappa < 1$, the diagonal correlation length
becomes very large $L^{(2)} \rightarrow \infty$ similarly
as the off--diagonal one. This is however, 
not necessarily the signature of strong correlations in the system.
In the considered case the magnitude of correlations vanishes
in large $N$ limit within the whole system, $g_2(x,0) \approx 0$ if
$|x| < \ell_1$. 
If attraction becomes very strong, $\kappa > 1$, then the diagonal
correlation length drastically decreases:
\begin{equation}
L^{(2)}_{diag} \approx 1/\sqrt{N^{\kappa-1}} = L^{(2)}_{off} \rightarrow 0.  
\end{equation}
In fact it becomes equal to the phase coherence length. 
Simultaneously the magnitude of these diagonal correlations  grows
to infinity, $g_2(x,0) \approx (N^{\kappa-1})^{d/2}$ if
$|x| < \ell_1$. 
This short range
diagonal order is related to the clustering of  particles. The 
joint probability
of finding two particles very close to each other is very large provided
that their separation is smaller than $L^{(2)}_{diag}$. Although the
particles are clustered the position of this `clustering spot' is
localized within the large distance determined by the spatial extension
of the center of mass ground state, $L^{(2)}_{diag} \ll \ell_1 \approx 1$.
This behavior is illustrated in Fig.4. The detailed comparison with
corresponding off-diagonal correlations, Fig.3 is very interesting. The 
presence off the off-diagonal long range order in the system is accompanied
by the absence of the diagonal order for relatively weak interactions,
$\kappa < 1$.  If interaction becomes very strong, $\kappa >1$, then
a diagonal order becomes very large on a very small distance. 
Simultaneously the off-diagonal order disappears.
Both characteristic
length scales for diagonal and off-diagonal correlations are the same.

For the repulsive forces the diagonal
order depends  very weakly on the interaction strength. 
The magnitude of diagonal
correlations tends  to zero with increasing particle number, $g_2(x,0)
\approx 2N^{-1}-3N^{\kappa} \rightarrow 0$ while the 
correlation length is large $L^{(2)}_{diag}=N /\sqrt{1-2N^{\kappa} 
+ 2 N^{-1}} \rightarrow \infty$. The change of the behavior of the
diagonal order with the interaction strength $\kappa$ is similar 
to the case of attractive forces (see Fig.3). However, as the correlations are
practically negligible a small diagonal order which appears when
$\kappa < -1$ is almost unvisible. In general we can say that 
similarly as in the case of attraction, there are no diagonal correlations
when the system exhibits the off-diagonal long range order.

\section{Conclusions}
In our paper we used exactly soluble many-particle model to illustrate 
a rigorous procedure of definig  the condensate phase at zero temperature. By
diagonalizing one-particle reduced density matrix we were able to study in
details the role of interactions on the condensate. If the
interaction strength becomes large $|\kappa| >1$ the condensate disappears even
when the  system is in its ground state.  In the oscillatory model studied here
this total depletion of the
condensate has a critical character in the thermodynamic limit.

The existence of nonvanishing order parameter is
accompanied by the off-diagonal long range order. The off-diagonal
order is present on all levels of the hierarchy of reduced $s$--particle
density matrices ($s/N \ll 1$) as long as $|\kappa| <1$. 
There is no diagonal long range order if the system 
exhibits the coherence (off-diagonal order).  

All these results have been obtained in the framework of a model. The model
has some unrealistic features, its advantege is that it is exactly soluble.
Therefore all the results presented in the paper are rigorous 
within the model.

Most of the results are, however, model-independent and in fact were 
discussed previously but without rigorous proof. Thus we provided another 
way of justifying them by showing that they can be proved for model
interactions.

In particular the existence of (at least) two diffeent length scales, the
one related to the external potential and the other resulting from the combined
effect of 
interparticle interactions is a generic feature of all interacting systems.
Moreover, as the first  scale does not depend on the number of particles
and interaction strength the second one does. The last scale is related 
to the spatial extention
of the relative degrees of freedom. Our model shows that
there is a coherence in the system if the 
two length scales are of the same order (different collective eigenmodes 
are not spatially separated). 
However, when these two scales become
drastically differrent there is a spatial separation of different
phases and off-diagonal long range order is destroyed.

In the case of attraction the effect of destruction of the condensate
relays on the fact that particles cluster in the region of the size much
smaller than the center of mass range. On the contrary,
for repulsive forces the condensate disapears when the amplitude  
of the center of mass zero-point
oscillations is much smaller than the size of the system.
In fact in the case of realistic attractive forces 
it is well known that interactions lead to clustering of atoms.   
Therefore condensates
with negative scattering length are limited to relatively small number
of atoms \cite{Dalfovo}.
Our results suggest however, that there is a similar limitation on the size of 
the condensate  interacting by repulsive forces. In fact in the 
case of short range forces the Thomas-Fermi approach gives the
ratio of the condensate size, $R$, to the characteristic length scale, 
$a_{ho}$, of the trap potential in the 
form: $R/a_{ho} = (15 N a_s/a_{ho})^{1/5}$
whre $a_s$ is the scattering length. For the largest existing trapped
condensates ($N \sim 10^7$) this value is about 10. This somehow  
fits perfectly to the scenario of destruction of the
condensate by interactions presented in this paper.   

\section*{Acknowledgement}
This work was supported by the KBN grant 2 P03B 13015.

\end{document}